\documentclass[aps,prb,showpacs,floatfix,twocolumn]{revtex4-1}
\usepackage{amssymb}
\usepackage{dcolumn}
\usepackage{amstext}
\usepackage{graphicx}
\usepackage{ulem}
\usepackage{cancel}
\usepackage{amsmath}
\usepackage{dsfont}
\usepackage[version=3]{mhchem}
\usepackage{color}

\newcommand{\ua}{\uparrow}
\newcommand{\da}{\downarrow}

\newcommand{\be}{\begin{equation}}
\newcommand{\ee}{\end{equation}}
\newcommand{\ba}{\begin{eqnarray}}
\newcommand{\ea}{\end{eqnarray}}

\newcommand{\etal}{\textit{et al.~}}

\begin{document}
\title{Local spin-density-wave order inside vortex cores in multiband superconductors  }
\author{Vivek Mishra}
\affiliation{Materials Science Division, Argonne National Laboratory, Lemont, IL
60439, USA}
\author{Alexei E. Koshelev}
\affiliation{Materials Science Division, Argonne National Laboratory, Lemont, IL
60439, USA}
 \date{\today}
\begin{abstract}
Coexistence of antiferromagnetic order with superconductivity in many families of
newly discovered iron-based superconductors has renewed interest to this old
problem. Due to competition between the two types of order, one can expect
appearance of the antiferromagnetism inside the cores of the vortices generated by the
external magnetic field. The structure of a vortex in type II superconductors holds
significant importance from the theoretical and the application points of view. Here
we consider the internal vortex structure  in a two-band s$_\pm$ superconductor near
a spin-density-wave instability.  We treat the problem in a completely
self-consistent manner within the quasiclassical Eilenberger formalism. We study the
structure of the s$_\pm$ superconducting order and magnetic field-induced
spin-density-wave order near an isolated vortex. We examine the effect of this
spin-density-wave state inside the vortex cores on the local density of states. 
\end{abstract}

\maketitle

\section{Introduction}
The emergence of superconductivity at the onset of magnetism is a hallmark of many
families of unconventional superconductors. Recently discovered iron based
superconductors (FeSCs) provided a new addition to this list. Parent compounds
for many of the FeSCs have the spin-density-wave (SDW) order, and
superconductivity (SC) appears upon doping or under pressure.
\cite{JohnstonReview,CanfieldReview,PJHReview,StewartReview,HHWenReview,ChubukovReview} 
The doping-temperature phase diagrams describing location of these two phases vary from 
material to material. In some systems, for example in the 1111 family \ce{RFeAsO$_{1-x}$F$_x$}  
(\ce{R} is a rare-earth element),  the SDW phase abruptly disappears once the SC
phase develops. In  other systems, such as 122 compounds based on
\ce{BaFe$_2$As$_2$}, the SDW order coexists with SC within some range of
parameters.

To understand how the SDW and SC  phases interact with each other and what triggers
superconductivity with such high transition temperatures, knowledge of the precise
structure of order parameters is essential. The local electronic structure near
defects can be extracted from scanning tunneling spectroscopy (STM) measurements and
it provides vital information about the order parameter. A Defect could be either  an
impurity or a topological singularity like a vortex induced by magnetic field. Here
we focus on structure of an isolated vortex.

It is well known that the shape of the vortex and the electronic structure close
to it are very sensitive to the gap structure.\cite{FischerReview} Copper oxide
based high-temperature superconductors have been subjected to extensive research
for various kinds of competing orders inside the vortex cores. The signatures of
such vortex-core orders have been reported in
\ce{Bi$_2$Sr$_2$CaCu$_2$O$_{8+\delta}$},\cite{Hoffman2002}
\ce{La$_{2-\delta}$Sr$_\delta$CuO$_4$},
\cite{Lake2001,Khaykovich2002,Lake2002,Tranquada2004,Khaykovich2005,Chang2007,Chang2008} 
\ce{YBa$_2$Cu$_3$O$_{7-\delta}$},\cite{Mitrovic2003,Haug2009,Wu2011,Chang2012,Leboeuf2013,Wu2013} 
\ce{YBa$_2$Cu$_4$O$_8$},\cite{Kakuyanagi2002} and
\ce{Tl$_2$Ba$_2$CuO$_{6+\delta}$}.\cite{Kakuyanagi2003} From the standpoint of
the theory, several different approaches have been adopted to explain these
experimental observations. Arovas \textit{et al.}\cite{Arovas1997} and Sachdev
\textit{et al.}\cite{Sachdev2004} studied antiferromagnetism in the vortex cores
within phenomenological Ginzburg-Landau free-energy functional method. Ghosal
\textit{et al.}\cite{Ghosal2002} used microscopic Bogoliubov-de Gennes (BdG)
technique to investigate this problem for the superconductors with a d-wave
symmetric order parameter. The BdG technique was used heavily by many
researchers to understand various aspects of the competing orders inside the
vortex cores.\cite{Zhu2001,Chen2005,Atkinson2008,Schmid2010} We also mention the
work of Garkusha \textit{et al.}\cite{Garkusha2005} where the Usadel equation
formalism has been used to explore the problem of antiferromagnetic vortex
cores. The Usadel equations, however, are only applicable in the dirty limit,
when the electronic mean free path is shorter than the coherence length, hence
only appropriate for s-wave superconductors.

The vortex state in FeSCs has been studied extensively by the
STM\cite{HoffmanReview} and several novel features near vortex cores have been
revealed. In the first study of the vortex structure in the optimally-doped
\ce{BaFe$_{1.8}$Co$_{0.2}$As$_2$} by Yin \textit{et al.}\cite{Yin2009} no subgap
states have been found. This is most probably due to large quasiparticle scattering
rate in this material.  On the other hand, optimally-doped
\ce{Ba$_{0.6}$K$_{0.4}$Fe$_2$As$_2$} does show peak at the vortex center, which is
shifted from the Fermi level to lower energy.\cite{Shan2011} This shift was
attributed to the quantum effect, which is realized in the materials with moderate values of
the product of the Fermi momentum $k_F$ and the coherence length $\xi_0$  at
temperatures lower than $T_c/( k_F \xi_0)$. Alternatively, such energy shift of the
localized state which breaks the particle-hole symmetry can be caused by magnetic
field-induced order in the vortex core. This scenario is very likely when a
superconductor is close to a SDW instability. This possibility, however, has not
been considered in Ref. \onlinecite{Shan2011}. Similar downshift was  found in
\ce{LiFeAs} by Hanaguri \textit{et al.},\cite{Hanaguri2012} even though this
material does not  have obvious proximity to magnetism. Song \textit{et
al.}\cite{Song2011}  studied the vortex state in \ce{FeSe} and found enhanced C$_4$
symmetry breaking in the vortex core which is probably related to orbital order in this material.

These compelling features have motivated many theoretical works. One class of
theories has associated the particle-hole asymmetric finite-energy peaks to the
normal-state  band structure of the materials.\cite{Araujo2009,Wang2010} In this
case the mechanism of particle-hole symmetry breaking is due to the quantum
effect discussed in Ref.\ \onlinecite{Hayashi1998}. Contrary to this proposal,
several other authors have considered orbital\cite{Hung2012},
nematic\cite{Chowdhury2011} or SDW order\cite{Jiang2009,Hu2009,Gao2011}. Hung
\etal\cite{Hung2012} have included orbital ordering within a self-consistent BdG
approach, and explained the enhanced C$_4$ symmetry breaking observed in FeSe by Song
\etal in Ref. \onlinecite{Song2011}. Similar results were reported by Jiang
\etal\cite{Jiang2009} and Hu \etal\cite{Hu2009} for the SDW order also using the
BdG method.

In this paper, we consider the emergence of the SDW order in the vortex cores
and its spectroscopic consequences. We use the quasiclassical Eilenberger approach
to study the problem of the field-induced SDW order inside an isolated vortex. Both the BdG and
Eilenberger approaches have their own advantages and complement each other. The BdG
method is more microscopic. On the other hand, the Eilenberger approach relies
on few most essential physical parameters. It is numerically less expensive and
allows to study more complex problems. %
We compute distribution of the superconducting and SDW order parameters inside
the core and typical length scales for both order parameters. We also
investigate influence of emerging SDW order on the density of states (DOS) near the
vortex. This paper is organized in the following manner. In the next Sec.
\ref{sec:model}, we describe the details of the model and the method. In Sec.
\ref{sec:results} we discuss the results and conclude in Sec.\ \ref{sec:conclusion}. 

\section{Model \& method}\label{sec:model}

\subsection{Quasiclassical equations for a two-band superconductor with spin-density wave}

\begin{figure}
\includegraphics[width=0.95\columnwidth]{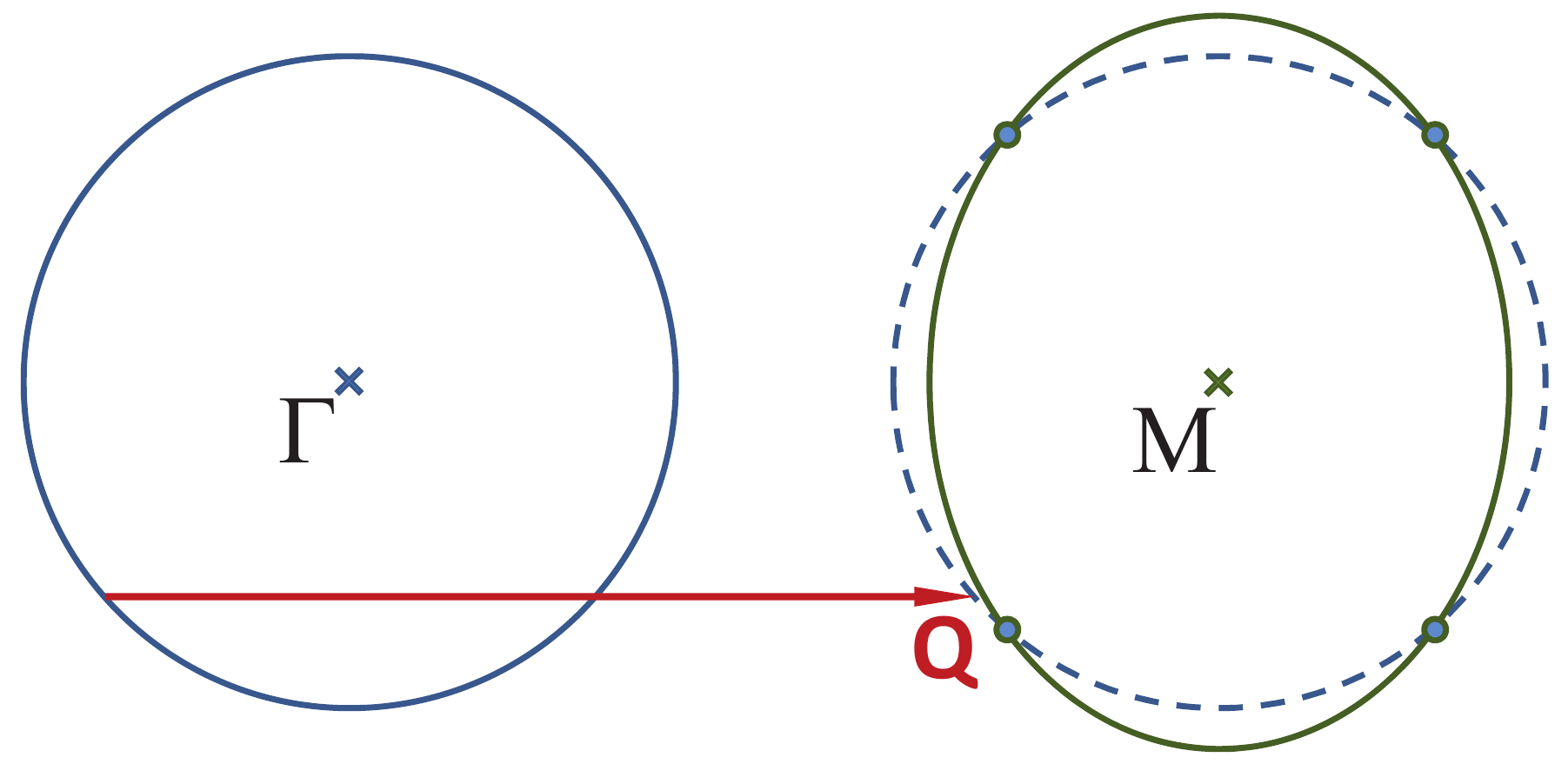}
\caption{(Color online) Schematic representation of
the holelike and electronlike Fermi surfaces (solid lines) centered around
the $\Gamma$ point and $M$ point respectively. A shifted holelike
Fermi surface is shown with dashed line. Filled circles are the hot spots,
where the nesting is perfect.}
\label{fig:fs}
\end{figure}

We consider a simple minimal model with two cylindrical Fermi surfaces, which allows us to
capture qualitative understanding of the problem. For the dispersion  of the
holelike Fermi surface, we take
\begin{eqnarray}
 \xi_h(\mathbf{k}) \equiv \xi_1(\mathbf{k})=\mu_h - \frac{k^2}{2m_h},
\end{eqnarray}
and for the electronlike Fermi surface we consider following dispersion,
\begin{eqnarray}
 \xi_e(\mathbf{k})\! \equiv\! \xi_2(\mathbf{k}\!-\!\mathbf{Q})\! =\!\frac{\left( k_x\!-\!Q_x\right)^2}{2m_e(1\!-\!\epsilon)}+\frac{\left( k_y\!-\!Q_y\right)^2}{2m_e(1\!+\!\epsilon)}\!-\!\mu_e,
\end{eqnarray}
where ($Q_x$,$Q_y$) is the SDW ordering vector and $\mu_h$,$\mu_e$ are the
energy offset for the hole and the electron band respectively. Fig. \ref{fig:fs} shows a schematic
picture of the two Fermi surfaces. It is useful to write these dispersions as,
\begin{eqnarray}
 \xi_h(\mathbf{k}) &=& -\xi, \\
 \xi_e(\mathbf{k}+\mathbf{Q}) &=& \xi+2\delta,
\end{eqnarray}
where $\delta$ is the energy scale, which  measures the deviation from perfect
nesting.
In general, $\delta$ is a function of the angle on the Fermi surface $\phi$ and
goes to zero at the hot spots (shown in Fig \ref{fig:fs}). For the dispersions
considered here,
\begin{equation}
 \delta(\phi) = \delta_{\text{iso}} + \delta_{\text{ani}} \cos 2\phi,
 \label{eq:nesting}
\end{equation}
with \begin{align*}
	\delta_{\text{iso}}  & =\frac{1}{2}\left(  \frac{m_{h}\mu_{h}}{m_{e}
		(1-\epsilon^{2})}-\mu_{e}\right),  \\
	\delta_{\text{ani}}  & =\frac{m_{h}\mu_{h}}{2m_{e}}\frac{\epsilon}
	{1-\epsilon^{2}},
\end{align*}
and  we treat $\delta_{\text{iso}}$ and $\delta_{\text{ani}}$ as tuning parameters.

The model Hamiltonian is same as used by several other
groups\cite{Vorontsov2009R,Vorontsov2010,Fernandes2010},
\begin{eqnarray}
 \mathbf{H}&=&\mathbf{H}_{\mathrm{kin}}+\mathbf{H}_{\mathrm{sc}}+\mathbf{H}_{\mathrm{sdw}}, \\ 
 \mathbf{H}_{\mathrm{kin}}&=& \sum_{i,\mathbf{k},\alpha} \xi_{i}(\mathbf{k})c^{\dagger}_{i,\mathbf{k},\alpha}c_{i,\mathbf{k},\alpha},\\ 
  \mathbf{H}_{\mathrm{sc}}&=& \sum_{i,\mathbf{k},\alpha,\beta} \left[ \Delta_{i} \left( \imath\sigma^{y}\right)_{\alpha\beta} c^{\dagger}_{i,\mathbf{k}, \alpha} c^{\dagger}_{i,-\mathbf{k}, \beta}
   + h. c. \right],\\ 
   \mathbf{H}_{\mathrm{sdw}}&=& \sum_{\mathbf{k},\alpha,\beta} \left[ M^{*} \left( \sigma^{z}\right)_{\alpha\beta} c^{\dagger}_{1,\mathbf{k}, \alpha} c_{2,\mathbf{k}, \beta} + h. c. \right], 
   \label{eq:Hamiltonian}
\end{eqnarray}
where $\Delta_{i}$ are the SC order parameters for two bands with $i=1,2$ being
the band index and $M$ is the SDW order parameter. We only consider singlet
superconductivity. For incommensurate SDW order $M$ is a complex quantity. Here
we consider only the commensurate SDW order, which makes $M$ a real quantity. We
will briefly discuss the consequences of incommensurability in the SDW order. 
The indices $\alpha$, $\beta$ denote the spin states, and
$c^\dagger_{i,\mathbf{k},\alpha}$ ($c_{i,\mathbf{k},\alpha}$) is the fermionic
creation (annihilation) operator for a fermion in the $i^{\mathrm{th}}$ band
with spin $\alpha$. We consider s$_\pm$ state for superconductivity with equal
gap magnitudes in two bands with a relative sign change.

The self-consistency conditions read,
\begin{eqnarray}
 \Delta_{i}&=&\sum_{j,\mathbf{k},\alpha,\beta} V^{\mathrm{sc}}_{ij}\left(-\imath\sigma^{y}\right)_{\alpha\beta}\left\langle c_{j,-\mathbf{k},\alpha}c_{j,\mathbf{k},\beta} \right\rangle, \\ 
 M&=& \sum_{k,\alpha,\beta} V^{\mathrm{sdw}} \left(\sigma^{z}\right)_{\alpha\beta}\left\langle c^{\dagger}_{1,\mathbf{k},\alpha}c_{2,\mathbf{k},\beta} \right\rangle.
\label{eq:OP_basic}
 \end{eqnarray}
Here $V^{\mathrm{sc}}$ and $V^{\mathrm{sdw}}$ are the pairing interactions for the
SC and SDW phases respectively and assumed to be momentum independent.

In the extended particle-hole basis,
$\Psi^{\dagger}=\left(c^{\dagger}_{1,\mathbf{k},\ua},c_{1,-\mathbf{k},\da},c^{\dagger}_{2,\mathbf{k},\ua},c_{2,-\mathbf{k},\da}\right)$,
the Hamiltonian reads,
\begin{eqnarray}
\mathbf{H}&=&\sum_{k}\Psi^{\dagger}\cdot \hat{\mathcal{H}}\cdot \Psi,\\
\hat{\mathcal{H}}&=&\left[\begin{array}{cccc}
\xi_1 & \Delta_1 & M & 0 \\
\Delta_1^* & -\xi_1 & 0 & M \\
M & 0 & \xi_2 & \Delta_2 \\
0&M&\Delta_2^* & -\xi_2\\
                     \end{array}\right].\label{eq:mfH}
\end{eqnarray}
The $4\times 4$ matrix Green's function for this mean-field Hamiltonian is
\begin{eqnarray}
 \hat{\mathcal{G}}=\left(i\omega \hat{\mathds{1}} - \hat{\mathcal{H}}\right)^{-1},
\end{eqnarray}
where $\omega=2\pi T(n+1/2)$ is the fermionic Matsubara frequency and $\hat{\mathds{1}}$ is 
4$\times$4 identity matrix in the two band particle-hole space.

Next, we derive the  quasiclassical equations, which were first obtained by
Eilenberger for conventional superconductors\cite{Eilenberger,Larkin1969}. These are
transportlike equations for the kinetic energy integrated Green's functions,
\begin{equation}
 \hat{g}=\frac{i}{\pi}\int d\xi\hat{\gamma} \cdot\hat{\mathcal{G}},
\label{eq:QCG}
\end{equation}
where $\hat{\gamma}$ is a 4$\times$4 diagonal matrix with elements ($1,-1,-1,1$). In
compact matrix form the quasiclassical equation reads,
\begin{align}
 &\left[\left(\omega+\frac{e}{ic}\mathbf{v}_F\cdot \mathbf{A}\right) \hat{\gamma}, \hat{g}\right]
 +\mathbf{v}_F\cdot \nabla\hat{g}\nonumber\\
 &+i\left[ (\hat{\mathcal{H}}_\delta+\hat{\mathcal{H}}_{\mathrm{sc}}+\hat{\mathcal{H}}_{\mathrm{sdw}})\hat{\gamma}, \hat{g}\right] =0,
 \label{eq:QC-g}
\end{align}
where $\mathbf{v}_F$ is the Fermi velocity and $\mathbf{A}$ is the vector potential.
$\hat{\mathcal{H}}_{\mathrm{sc}}$, $\hat{\mathcal{H}}_{\mathrm{sdw}}$ are the SC and the SDW
components of the mean-field Hamiltonian in the basis spanned by $\Psi$.
$\hat{\mathcal{H}}_\delta$ is a 4$\times$4 diagonal matrix with elements
(0,0,2$\delta$,-2$\delta$) containing information about nesting between the Fermi
surfaces. Its contribution drops out in the absence of the SDW order. 
\begin{figure}
\includegraphics[width=0.9\columnwidth]{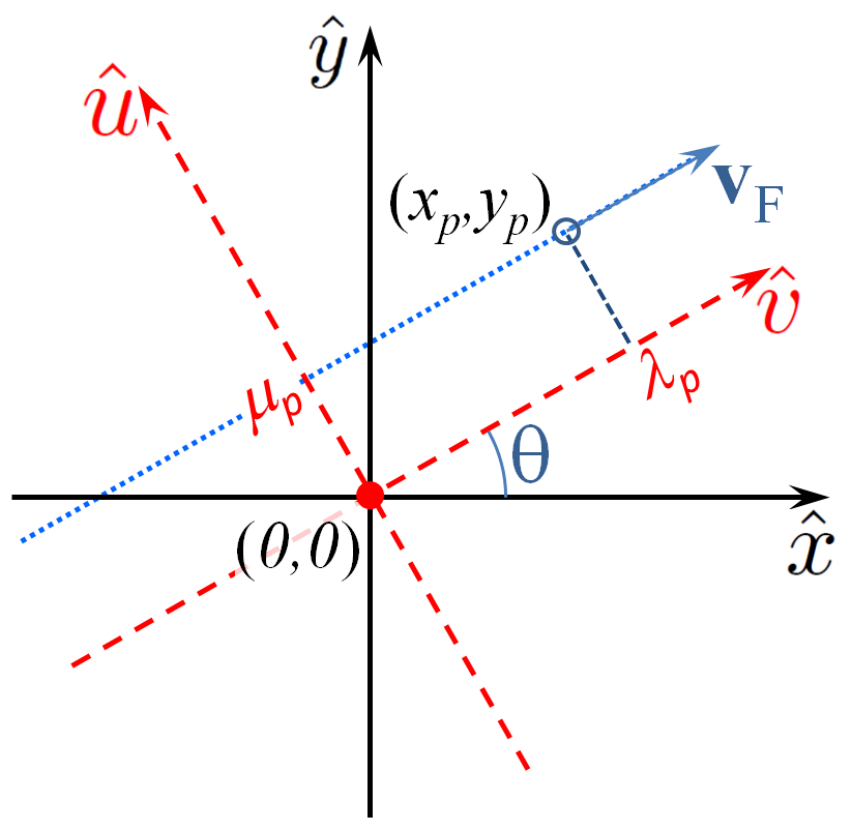}
\caption{(Color online) Coordinate systems used to solve the Eilenberger
equations. The real space lab frame is shown with solid lines, while the dashed
lines represent the new coordinate system. Filled circle is the location
of the vortex core and chosen as the origin. Open circle is the point where the solution is
required. A point in the lab frame $\textit{r}=x_p \hat{x} + y_p \hat{y}$ maps
to $\lambda_p \hat{v} + \mu_p \hat{u}$ in the new coordinate system.}
\label{fig:cs}
\end{figure}
This equation agrees with one derived by Moor \etal\cite{Moor2011}.
In the pure superconducting
limit Eq.\ (\ref{eq:QC-g}) reduces to the well-known Eilenberger equation,
\begin{equation}
\mathbf{v}_F\cdot \nabla \hat{g}\!+\!\left[\left(
\omega\!+\!\frac{e}{i c}\mathbf{v}_F \cdot \mathbf{A}\right) \hat{\gamma}, \hat{g}\right]
\!+\!i\left[ \hat{\mathcal{H}}_{\mathrm{sc}}\hat{\gamma}, \hat{g}\right]\!=\!0.
\end{equation}
Equation (\ref{eq:QC-g}) has to be supplemented with the normalization condition,
\begin{equation}
\hat{g}^2=\hat{g}^2_{bulk}.
\end{equation}
In particular, $\hat{g}^2_{bulk}=\hat{\mathds{1}}$ for a uniform superconductor without SDW
order.
The self-consistency conditions for the order parameters can be expressed in terms
of the Eilenberger functions as,
\begin{eqnarray}
 \Delta_1 &=& i \pi T \sum_{\omega} \left\langle V^{\mathrm{sc}}_{11} g_{12} + V^{\mathrm{sc}}_{12} g_{34} \right\rangle_{F.S.}, \\
 \Delta_2 &=& i \pi T \sum_{\omega} \left\langle V^{\mathrm{sc}}_{12} g_{12} + V^{\mathrm{sc}}_{22} g_{34} \right\rangle_{F.S.}, \\
 M &=&  i \pi T \sum_{\omega} \frac{V^{\mathrm{sdw}}}{2}\left\langle g_{13}+g_{24} \right\rangle_{F.S.}.
\end{eqnarray}
Here $g_{ij}$ are components of $4\times 4$ matrix Green's function (indices
$i,j\!=\!(1,2)$/$(3,4)$ correspond to the hole/electron band). $\left\langle
g_{ij}\right\rangle_{F.S.}$ means angular average over the Fermi surface for the
respective bands weighted with the density of states, which is approximately the
same for the both bands for the Fermi surfaces considered here.  As for the
pairing-interaction matrix, to get  the s$_\pm$ superconducting state we take into account
only the interband repulsive interaction and neglect intraband terms, i.e.,
$V^{\mathrm{sc}}_{11}=V^{\mathrm{sc}}_{22}=0$. For pure SC state, Eq. (17)
equations can be transformed to a set of Riccati equations, which makes the
numerical solution much easier.\cite{Schopohl1998} However this transformation
is not useful for the problem considered here. In the following section, we
discuss the strategy to solve these equations for the present case.

\subsection{Numerical solution}

The Eilenberger equations are first-order partial differential equations. A
standard tool for solution of this type of equations is the method of
characteristics. The basic idea of this method is to introduce the new
coordinate system, in which the partial differential equation reduces to an
ordinary differential equation. Hence, it is useful to introduce a coordinate
system spanned by the two orthogonal vectors, a unit vector along the direction
of the Fermi velocity $\hat{v}$ and  a unit vector $\hat{u}$ orthogonal to
$\hat{v}$, see Fig. \ref{fig:cs}.  The unit vectors spanning the new coordinate
system read,
\begin{eqnarray}
 \hat{v} &=& \cos \theta \hat{x} + \sin \theta \hat{y}, \\
 \hat{u} &=& -\sin \theta \hat{x} + \cos \theta \hat{y}.
\end{eqnarray}
Here $\theta$ is the angle between the Fermi velocity and the $x$ axis in the lab frame. A point in
the lab frame $\mathbf{r}=(x,y)$ maps to ($\lambda,\mu$) in the $\hat{v}$-$\hat{u}$
frame. A point ($x_p, y_p$) at which solution is desired transforms to
($\lambda_p,\mu_p$) as,
\begin{eqnarray}
 \lambda_p &=& x_p \cos \theta + y_p \sin \theta, \\
 \mu_p &=& -x_p \sin \theta + y_p \cos \theta,
\end{eqnarray}
where the parameter $\mu_p$ has the meaning of an impact parameter. For a fixed trajectory,
this impact parameter is uniquely determined by ($x_p,y_p$) and it does not change
with change of $\lambda$. The quasiclassical equations are solved along these
classical trajectories ($\hat{v}$) in the real space. Along such trajectories
quasiclassical equations reduce to system of ordinary differential equations, which
are much easier to handle than solving a set of partial differential equations.
\begin{figure}
\includegraphics[width=0.95\columnwidth]{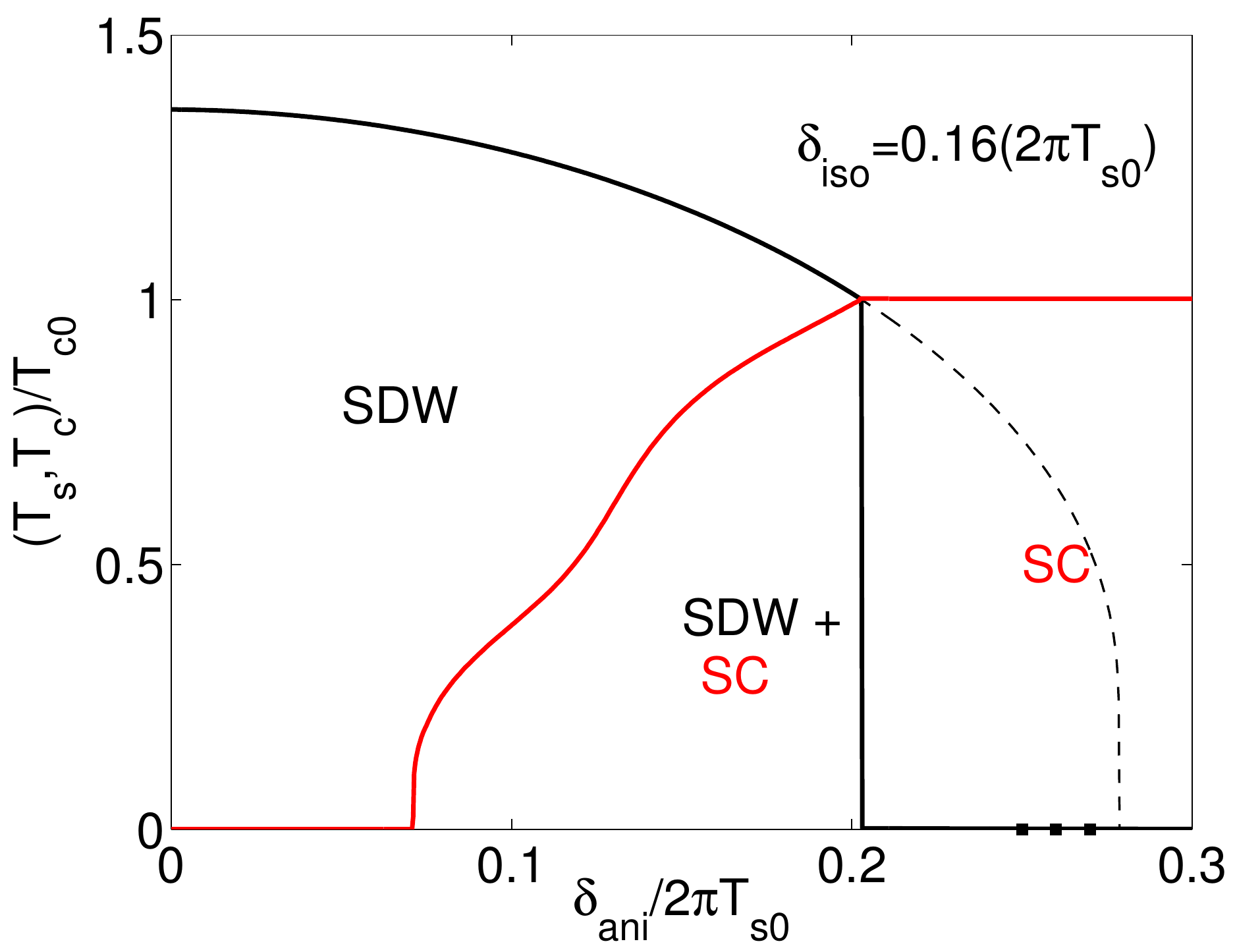}
\caption{(Color online) The phase diagram
in the (T,$\delta_{\text{ani}}$) plane for the two phases at zero magnetic field. Thick black/red line
indicates transition to the SDW/SC state. The thin dashed line shows the $T_s$ in the absence of the
SC correlations. Filled squares are considered as representative cases
in this paper.}
\label{fig:phase_diagram}
\end{figure}
Far away from the bulk, the system is homogeneous. The homogeneous
values are used as initial values. Now on a given trajectory, there are two possibilities.
One can integrate towards the defect (vortex in this case) from the two extreme ends ($\lambda=\pm \infty$)
of the trajectory. Due to the first-order nature of the equations, the numerical solution
readily converges to exponentially growing function. Of course, these exponentially growing solutions
are unphysical. However, it is possible to construct the physically bounded solution
at any point using the exponentially growing solution using the explosion method.
The explosion method exploits exponentially growing solutions to obtain the physical
solution.\cite{Thuneberg1984,Thuneberg1987,Klein1987} (See Appendix \ref{app:em} for
details)
For each point in the real space, one has to solve the Eilenberger equations
for all the trajectories and for each Matsubara frequency.
To obtain a physical solution, we solve the Eilenberger equations from two opposite directions
$\lambda=\pm \infty$ towards the point, where the solution is desired.
 As shown in the Appendix, the two exploding solutions $\hat{g}_\pm$ diverging in the $\pm \infty$ limits, provide
 the physical solution $\hat{g}_p$ as
 \begin{equation}
  \hat{g}_p = \frac{\hat{g}_- \hat{g}_+ - \hat{g}_+ \hat{g}_-}{\hat{g}_- \hat{g}_+ + \hat{g}_+ \hat{g}_-}.
 \end{equation}
Once all the Eilenberger functions are computed for an initial guess for the
order parameters, an updated set of order parameters is recalculated, and this
process continues till it converges to a solution. It is convenient to normalize all
the energy scales to $T_c$ and  all the lengths are measured in the unit of
superconducting coherence length  $\xi_0=v_F/(2\pi T_c )$. {Here $v_F$ is the 
average Fermi velocity of the two bands. We consider weak ellipticity for the electronlike 
Fermi surface, and the Fermi velocities of the two bands are roughly equal.} 
All our results are presented in these units.
\begin{figure}
\includegraphics[width=0.95\columnwidth]{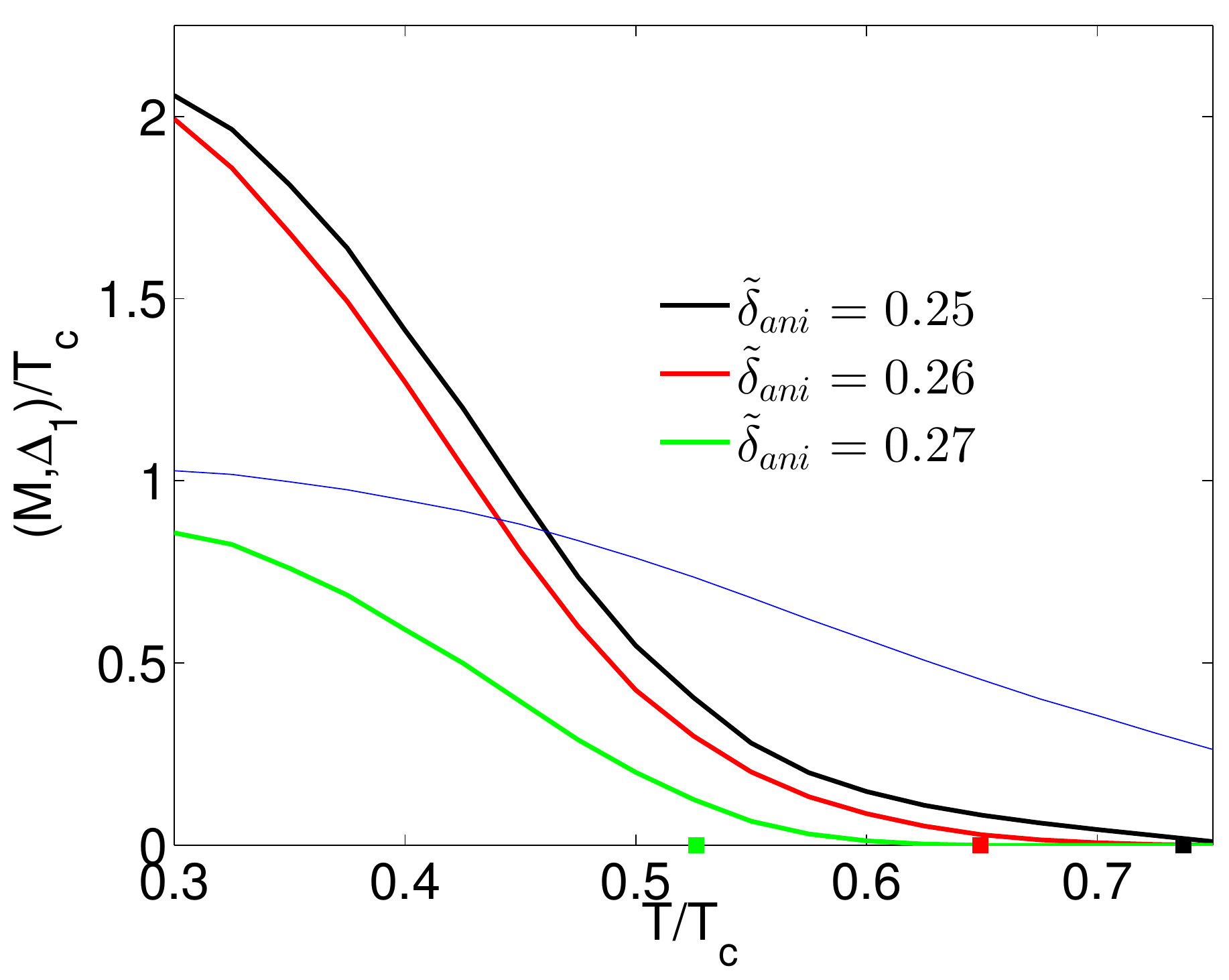} \caption{(Color online)
	The temperature dependence of the bulk superconducting order parameter and the
	SDW orders at the center of the vortex for three values of the parameter
	$\delta_{\text{ani}}/2\pi T_{s0}$= $0.25$,$ 0.26$ and $0.27$. All the energy
	scales are normalized to $T_c$.
	The thin line shows the temperature dependence of the magnitude of the SC order
	in the bulk, which is almost identical for two bands. This signs of SC order
	parameters are opposite for two bands. The mean-field SDW transition
	temperature in the absence of superconductivity is indicated with filled square
	for each case.} \label{gapT}
\end{figure}
\begin{figure}
\includegraphics[width=7.5cm, height=6.45cm]{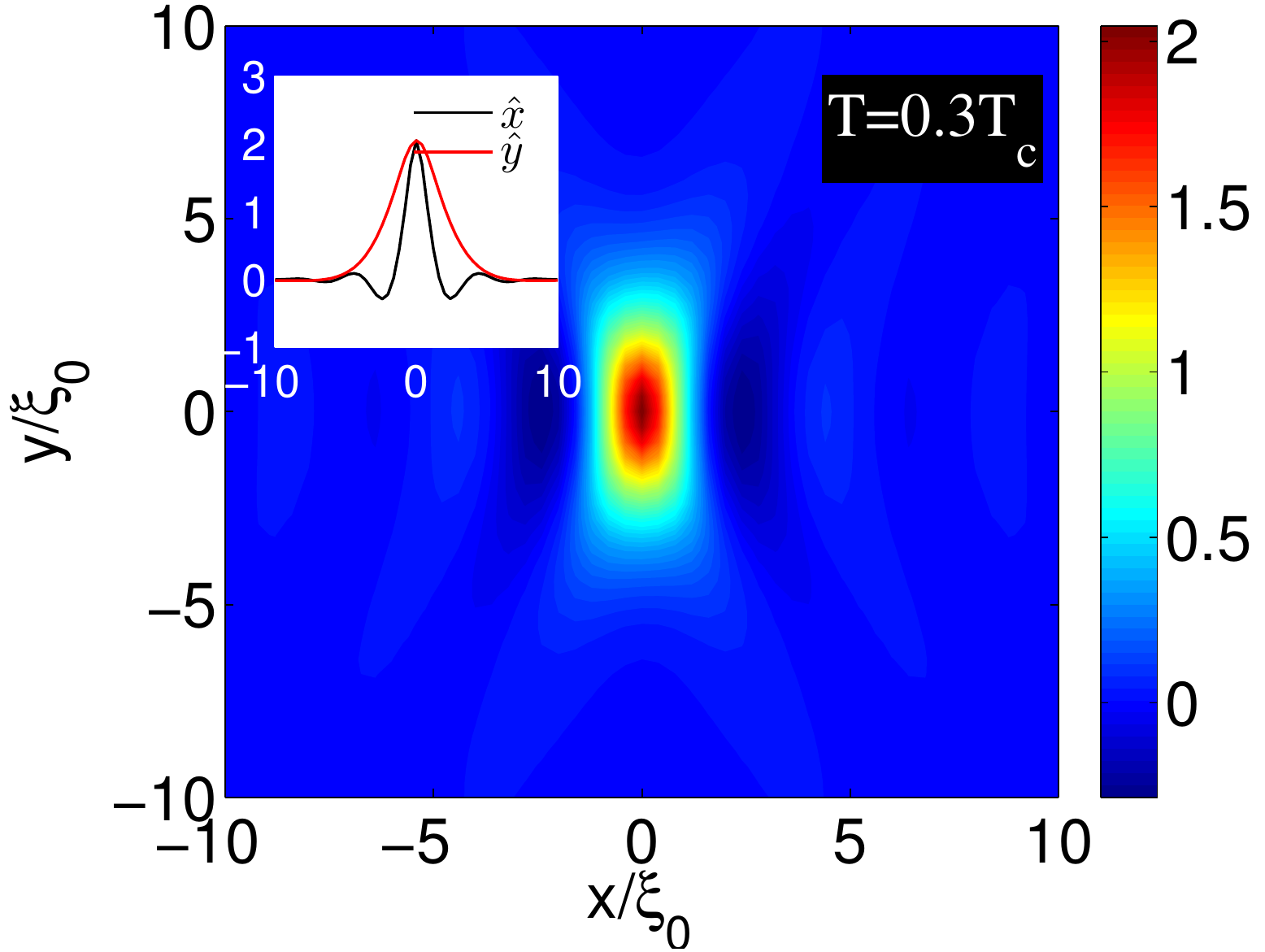}
\caption{(Color online) The spatial variation
of the magnitude of the SDW order at $T=0.3T_c$ for $\delta_{\text{ani}}=0.25$(2$\pi T_{s0}$),
where $T_{s0}$ is the SDW transition temperature for
a system with perfect nesting and with the same interaction strength. The magnitude of the SDW order
is normalized to $T_c$. The hot spots are located near the y-axis.}
\label{sdwvortex}
\end{figure}
\begin{figure*}
\includegraphics[width=7.5cm, height=6.5cm]{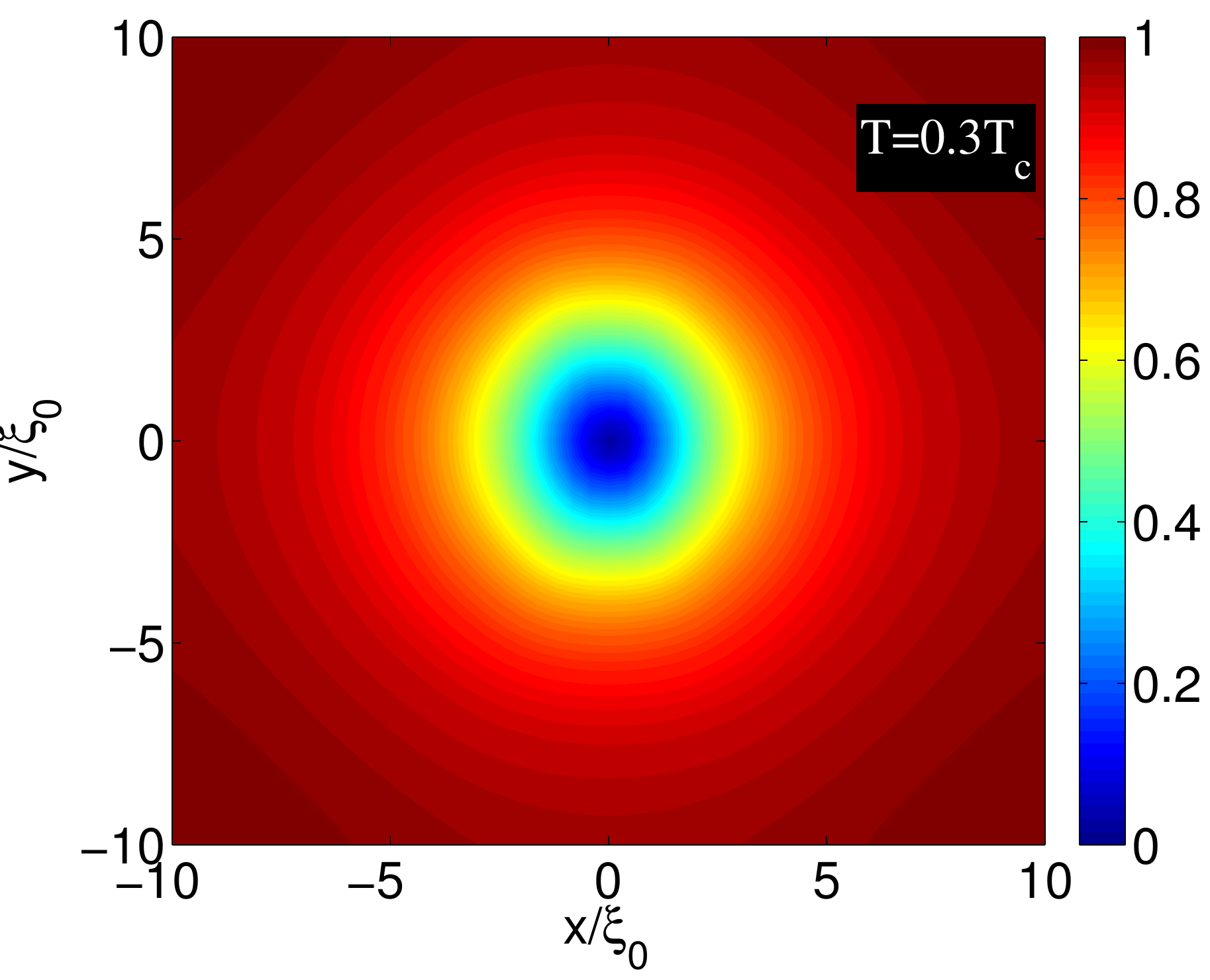}
\includegraphics[width=7.5cm, height=6.5cm]{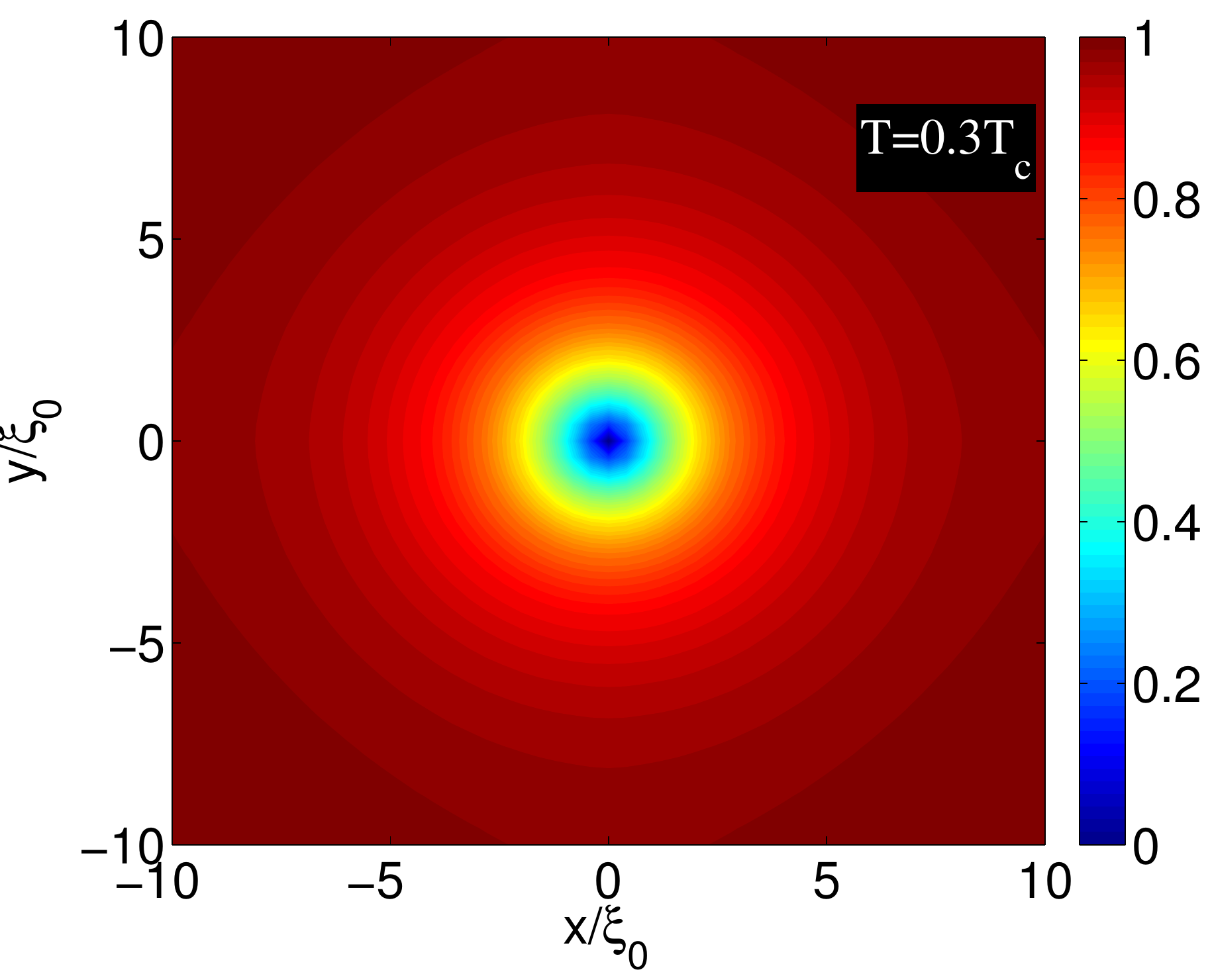}
\caption{(Color online)  The left panel shows the gap magnitude normalized to its bulk
value at $T=0.3T_c$ in the presence of the SDW order and the right panel shows the
gap structure at the same temperature with no SDW order for the hole band.} \label{vortex}
\end{figure*}

\section{Results \& Discussion}\label{sec:results}

Coexistence of the SDW state and the superconductivity is very sensitive to the
underlying electronic structure. For the two-band model we consider, the nesting
function $\delta(\phi)$ in Eq.\ (\ref{eq:nesting}) can be tuned to get a
co-existing phase.\cite{Vorontsov2009R,Vorontsov2010,Fernandes2010} Here we are
interested in a situation where there is no long-range SDW order in the absence
of the magnetic field. Fig. \ref{fig:phase_diagram} shows the phase diagram as a
function of the anisotropic nesting parameter $\delta_{\text{ani}}$ for a fixed value
of $\delta_{\text{iso}}=0.16(2\pi T_{s0})$. The presence of the superconductivity strongly modifies
the SDW phase boundary (thin dashed line in Fig. \ref{fig:phase_diagram}). The
region between the original and SC-renormalized phase boundaries
provides  a possibility of the SDW order in the regions, where the SC phase get
suppressed locally. The phase diagram shown in Fig. \ref{fig:phase_diagram} is
only includes the commensurate SDW phase. Vorontsov \etal \cite{Vorontsov2010}
have shown that the incommensurate SDW phase may co-exist with the SC in  a
larger area of the phase diagram. We consider few representative cases with
$\delta_{\text{ani}}/2\pi T_{s0}= 0.25,0.26$ and $0.27$. Here $T_{s0}$ is the
SDW transition temperature for a system with perfect nesting and with the same
interaction strength. We have set $T_{s0}=2T_{c0}$, where $T_{c0}$ is the
superconducting temperature. As we are mostly interested in low-temperature
behavior, we restrict our calculations below $T=0.8T_c$.

\subsection{Order parameter profiles}

Fig. \ref{gapT} shows the temperature dependence of the bulk superconducting gap
and the SDW order in the core. It is evident that the SDW order appears roughly
below the temperature one would expect from the phase diagram shown in Fig.\
\ref{fig:phase_diagram}. The temperature dependence of the SDW order parameter
deviates strongly from the mean-field behavior ($\propto \sqrt{1-T/T_c}$). We
see two different temperature regimes. At lower temperature the SDW order grows
strongly, but slightly below the mean-field SDW transition $T_s$ it develops a
tail, which survives even above $T_s$. It should be noted here that the phase
diagram is based on the commensurate SDW phase for the normal state electronic
structure. However, the electronic structure of the vortex core states is not
the same as in the normal state. The onset of the SDW order is mostly determined
by the core bound states. The SDW transition temperature T$_{s}\propto
\exp[-1/N_0 V_{\mathrm{sdw}} f(\delta_{\mathrm{iso}},\delta_{\mathrm{ani}})]$,
where the function $f(\delta_{\mathrm{iso}},\delta_{\mathrm{ani}})$ depends on
nesting parameters, $V_{\mathrm{sdw}}$ is the SDW interaction and $N_0$ is the
density of states. As in the vortex core the density of states is higher than
the normal-state value, the SDW onset temperature may exceed the mean-field
value.

\begin{figure}
\includegraphics[width=.9\columnwidth]{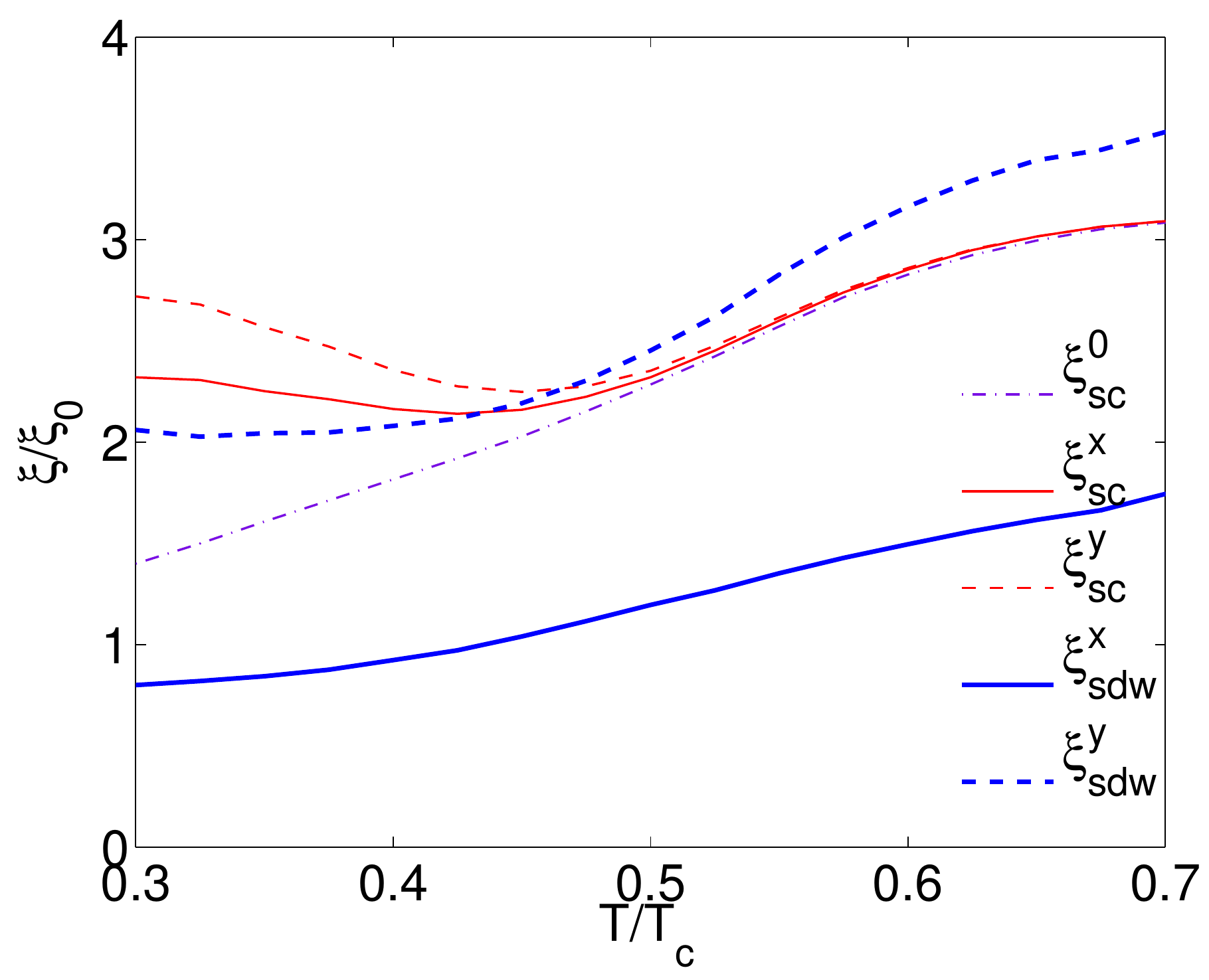}
\caption{(Color online) The temperature dependence of
characteristic length scale for SDW order ($\xi^{\mathrm{sdw}}_{x/y}$) and  the
characteristic length scale of the SC order ($\xi^{\mathrm{sc}}_{x/y}$) on the hole band for
$\delta_{\text{ani}}/2\pi T_{s0}= 0.25$. Subscripts denotes
$x$ and $y$ directions in the real space. The same characteristic length
scale for a pure superconductor is plotted with a dotted dashed line for comparison.
}
\label{lengths}
\end{figure}

We restrict ourselves to the commensurate case only, but for the incommensurate
case the phase boundary shifts towards slightly higher temperatures. As reported
by Vorontsov \etal in Ref. \onlinecite{Vorontsov2010}, an incommensurate order
may exist in a larger portion of the phase diagram. We have also performed
calculations, where we allow incommensurability in the SDW order. With
incommensurability  the SDW order parameter becomes complex and acquires a
finite phase. We found that the phase of the SDW order parameter is temperature
dependent and varies very weakly in the real space. The incommensurate order
persists above the phase boundary shown in Fig. \ref{fig:phase_diagram}. Since
we did not find anything qualitatively different, we will focus on the
commensurate case only. Furthermore, there is no qualitative difference between
the cases considered here, except for the temperature dependence.  Therefore, we
continue our discussion with $\delta_{\text{ani}}/2\pi T_{s0}= 0.25$. For this
$\delta_{\text{ani}}$ and $\delta_{\text{iso}}/2\pi T_{s0}= 0.16$ the nesting
hot-spot angle $\phi\approx 64.9^{\circ}$ is close to the $y$ direction which
strongly influences anisotropic properties of the vortex. Figure \ref{sdwvortex}
shows the magnetic field-induced SDW order parameter in the real space at
$T=0.3T_c$. Spatial coordinates have been normalized to the superconducting
coherence length $\xi_0$ and the SDW order is normalized to T$_c$. An important
feature is the oscillations of the  SDW order along the $x$ direction which is 
most clearly seen in the inset. At lower temperatures when the SC vortex is very
small, the SDW state is localized very close to the vortex core. As the
temperature grows, vortex becomes larger and the region with the SDW order also
increases, due to a larger region of the suppressed superconductivity. The
size of the vortex is larger in the presence of the field-induced SDW order. This
property can be seen in Fig. \ref{vortex}, in which we compare a SC
vortex with and without the magnetic field-induced SDW order. The SDW order
makes vortex larger and anisotropic. The intrinsic anisotropy of the underlying
band structure is weak. Hence the large anisotropy in the real space is mainly
due to the field-induced SDW order. Strong enhancement of the anisotropy is
reflected in the characteristic length scales associated with the SC order along
the two principal directions. Fig. \ref{lengths} shows the length scales
associated with the SDW ($\xi^{\mathrm{sdw}}_{x/y}$) and the SC order
($\xi^{\mathrm{sc}}_{x/y}$) along the $x$ and $y$ axis. We define the
superconducting coherence length $\xi_{\mathrm{sc}}$ as a distance from the
vortex core, where the order parameters reaches half of its bulk value.
Similarly, the magnitude of the SDW order drops to half of its value at the core
at a distance $\xi_{\mathrm{sdw}}$ from the vortex core. As illustrated in Fig.
\ref{lengths}, the SC length scales along $x$ and $y$ directions become
different in the presence of the SDW order and this is mainly due to anisotropy
in the field-induced SDW order. Note that the SDW correlations are stronger
along the $y$ direction which is closer to the nesting hot spots.  This causes
stronger suppression of the SC order and reduces the SC characteristic length
along this direction. Next, we discuss the density of states near the vortex
core.
\begin{figure*}
\includegraphics[width=0.98\columnwidth]{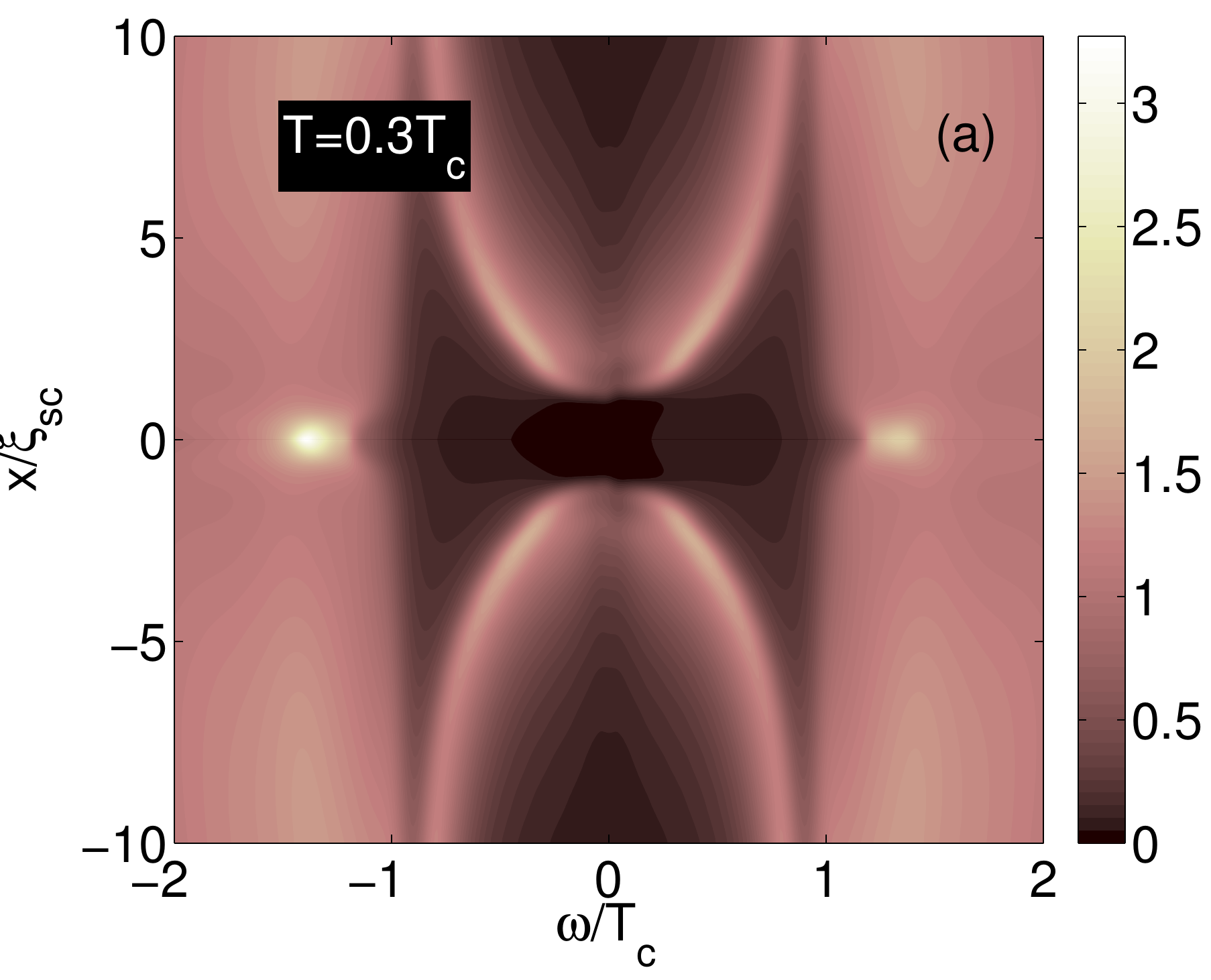}
\includegraphics[width=0.98\columnwidth]{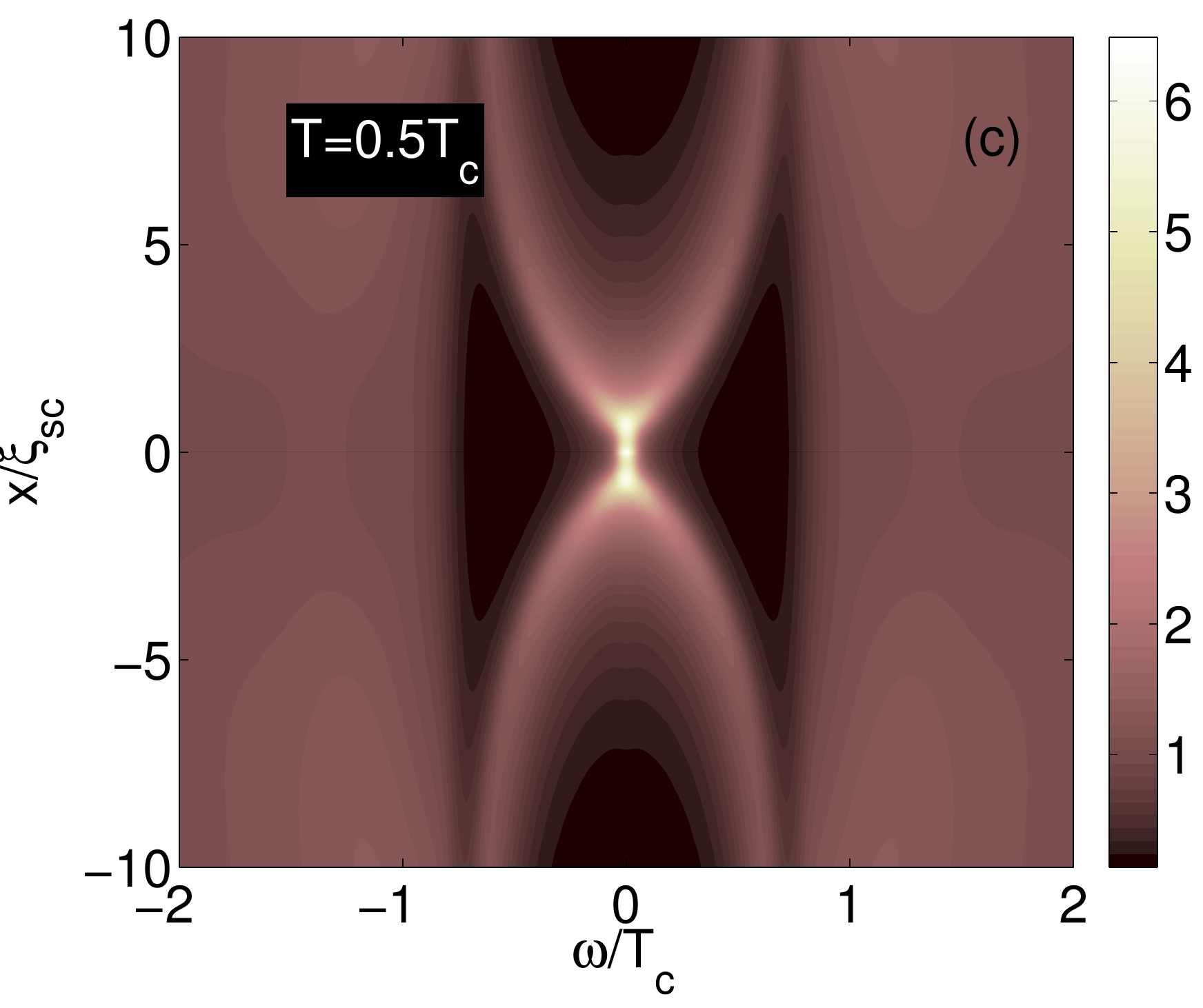}
\includegraphics[width=0.98\columnwidth]{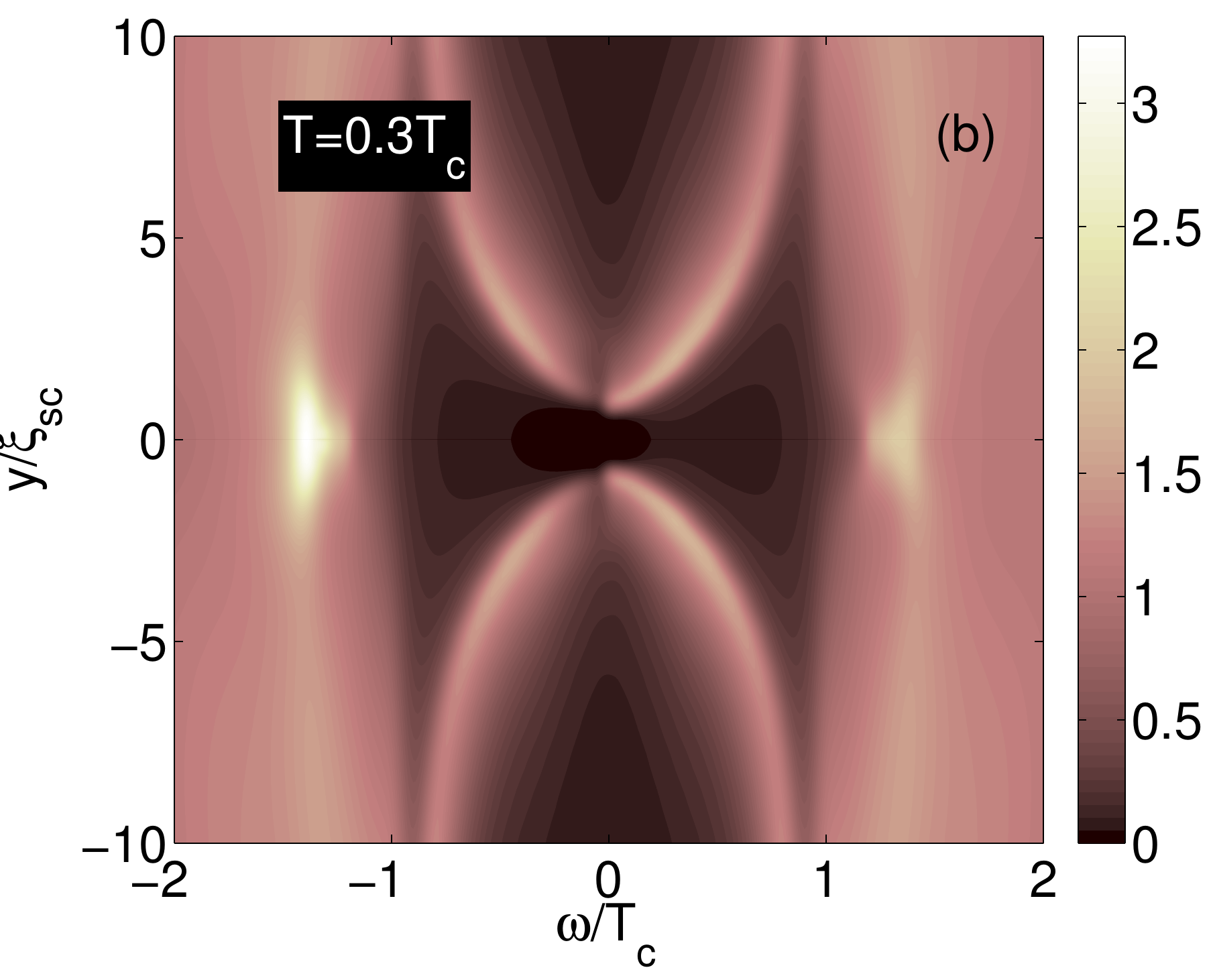}
\includegraphics[width=0.98\columnwidth]{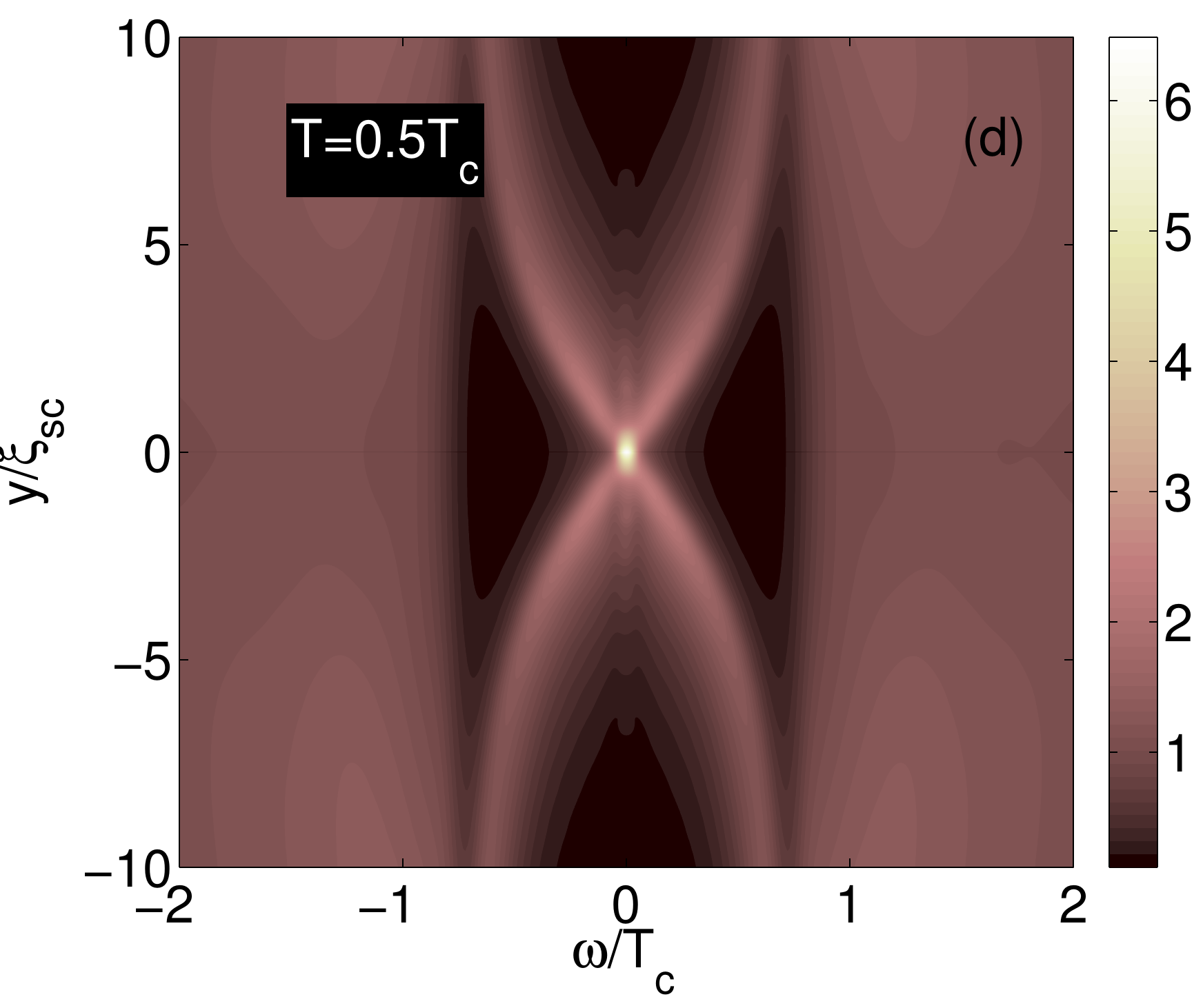}
\caption{(Color online)  DOS along the $x$  direction (first row) and along $y$ direction (second row) at
T=0.3$T_c$ in panel (a), (b) and at T=0.5$T_c$ in panel (c) and (d) for $\delta_{ani}/2\pi T_c$=0.26.} \label{vortex_dosx}
\end{figure*}
\begin{figure*}
\includegraphics[width=0.98\columnwidth]{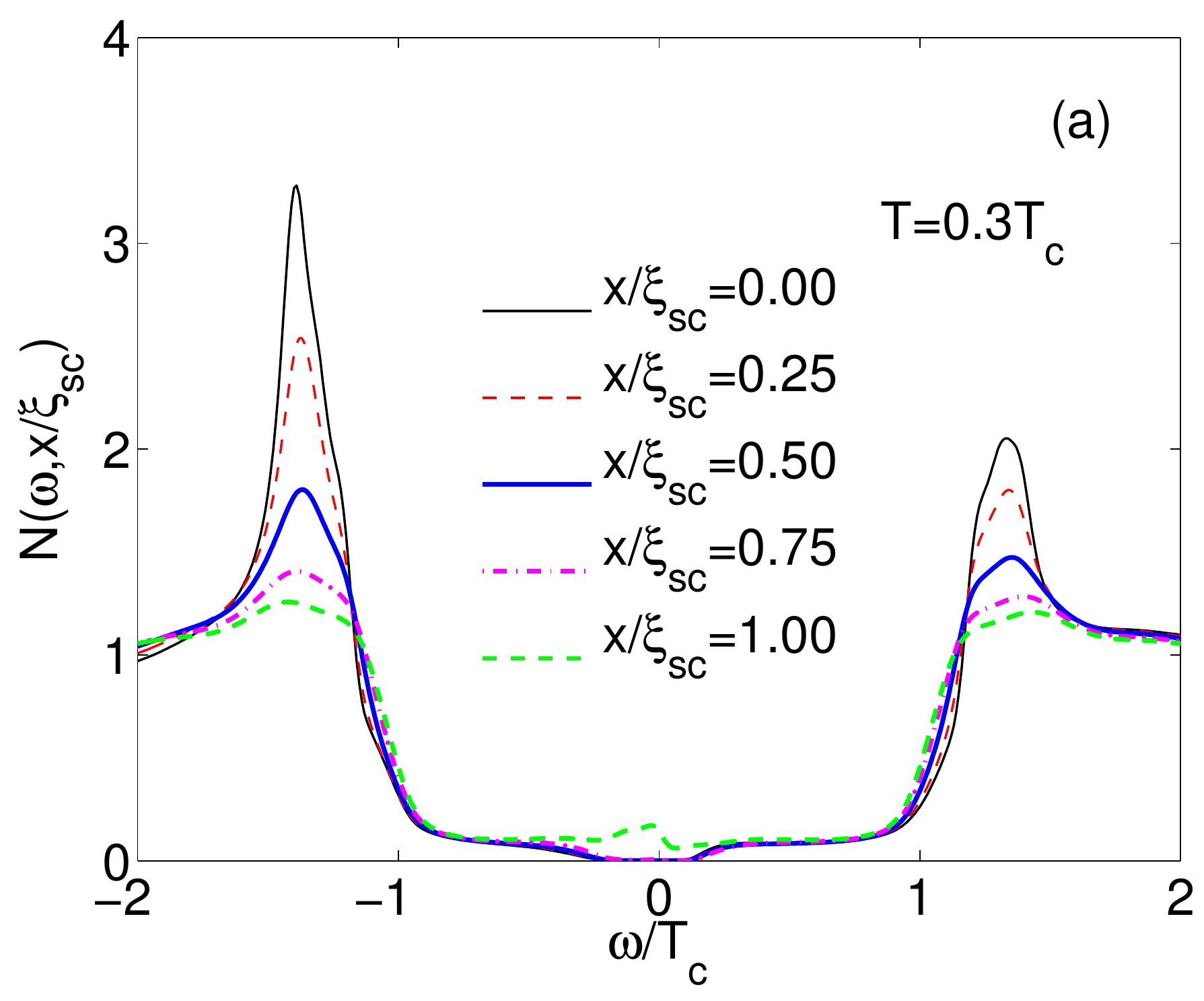}
\includegraphics[width=0.98\columnwidth]{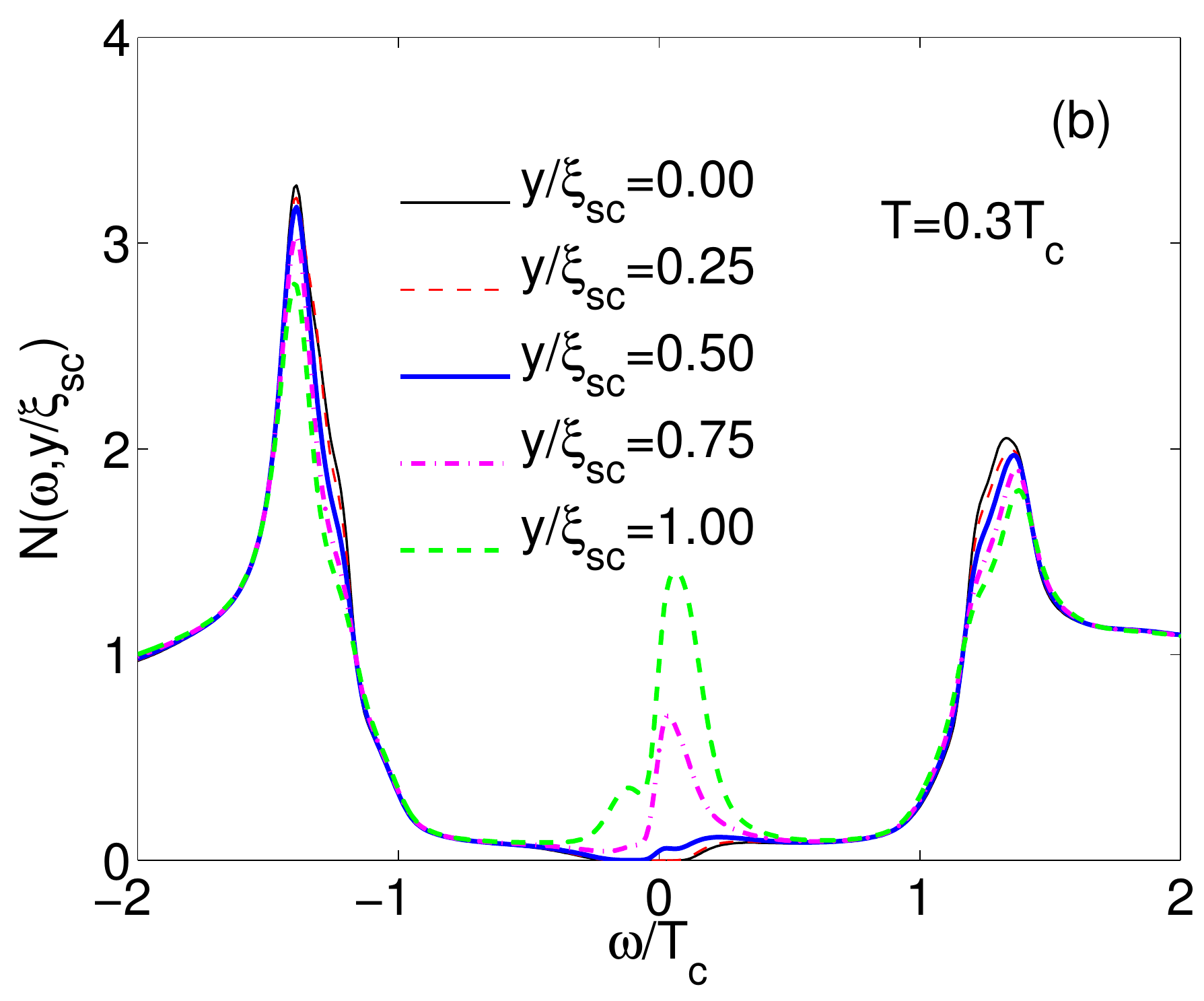}
\caption{(Color online) Panels (a) and (b) show DOS near the vortex core along
$x$ and $y$ directions respectively at T=0.3$T_c$.} \label{vortex_doscut}
\end{figure*}

\subsection{Density of states} All the previous calculations were done using the
Matsubara frequencies. For the DOS calculation it is necessary to go to the real
frequencies. We used the order parameter profiles calculated in the Matsubara
representation. The analytic continuation, $i\omega\rightarrow E+i\eta$, is done
with an artificial broadening $\eta=0.05T_c$. 
This process also requires solution of the
Eilenberger equations for the real frequencies using the same approach as in the
calculation of the order parameters in the previous section. The total DOS is
the sum of the partial DOSs for each band and, in terms of the Eilenberger functions,
it is given by
\begin{equation}
N(E)=\mathrm{Re}\langle[g_{11}(E+i\eta)+g_{33}(E+i\eta)]\rangle_{FS}.
\end{equation}
Fig. \ref{vortex_dosx} presents evolutions of the density of states along the
two principal directions for two temperatures. The first row of Fig.
\ref{vortex_dosx} shows DOS along the $x$ axis, which is away from the hot spots
and the second row shows DOS along the $y$ axis which is closer to the hot
spots. The field-induced SDW order enhances the violation of the C$_4$
rotational symmetry the vortex center. For conventional superconductors, the DOS
is always particle-hole symmetric, unless the superconductor is in the quantum
regime when $k_F\xi_0$ is not too large and $T\ll
T_c/k_F\xi_0$.\cite{Hayashi1998}  Another key feature of the classical
clean-limit DOS in the vortex core is the sharp peak at zero energy
corresponding to the localized state. The quantum effects shift this zero-bias
peak to the finite energy but, do not break the rotational symmetry. The
emergence of the SDW order leads to particle-hole asymmetry in the DOS and also
strongly violates the C$_4$ symmetry. The particle-hole asymmetry is bigger
along $x$ direction because of stronger deviation from nesting in this
direction. Another important feature which is visible in the Fig.\
\ref{vortex_dosx} is suppressed spectral weight at the core indicating that the
energy of localized state is shifted from zero to a finite value corresponding
to opening of a minigap in the core. This is shown more clearly in Fig.
\ref{vortex_doscut} in which we show the DOS plots at several representative
points.  This small gap in the DOS near the vortex core is particle-hole
asymmetric, which is a hallmark of energy gap due to the SDW order. This gap
vanishes away from the vortex core indicating presence of a state with energy close
to zero localized outside the core. As the temperature increases and the SDW
order weakens, the apparent gap in the core disappears as shown in panel (c) and
(d) of Fig. \ref{vortex_dosx} at T = 0.5$T_c$. The described features are the keys to
distinguish between the quantum effect and the field-induced SDW order. Figure
\ref{vortex_dosx} also shows the DOS for two different temperatures. The
temperature dependence of the DOS is easy to understand. As the temperature
increases, the SDW order weakens, which reduces the degree of C$_4$ symmetry
breaking and the particle-hole asymmetry in the DOS.

\section{Summary and conclusion}\label{sec:conclusion} 

We study the structure of an isolated superconducting vortex near a SDW
instability inside the superconducting dome. We show that the SDW order develops
inside the vortex below a critical temperature determined by the strength of the
SDW instability. This leads to C$_4$ symmetry breaking near the vortex core. If
there is already C$_4$ breaking in underlying band structure, then it gets
strongly enhanced due to the SDW order near the vortex core. The corresponding
deformation of the vortex shape can be imaged by the STM technique. We find that
the field-induced SDW order persists beyond the superconducting vortex region.
The tunneling DOS carries very strong signatures of this field-induced order. A
small energy gap develops inside the core and gives rise to strong particle hole
asymmetry, which is pronounced along the directions away from the hot spots. Our
results are in qualitative agreement with the STM data on on
\ce{Ba$_{0.6}$K$_{0.4}$Fe$_2$As$_2$}\cite{Shan2011} which may indicate the
presence of the vortex-core SDW order in this material. Our findings also agree with
BdG-based works by other groups. 
\begin{acknowledgments}
This work was supported by the Center for Emergent Superconductivity, an Energy
Frontier Research Center funded by the US DOE, Office of Science, under Award No.
DE-AC0298CH1088.
\end{acknowledgments}

\appendix
\section{Explosion method}\label{app:em}

The general structure of the Eilenberger equations is,
\begin{equation}
 \frac{d\hat{g}}{d\lambda} = \left[ \mathcal{T},\hat{g}\right],
\end{equation}
which is a first-order ordinary differential equation. It is straightforward to show
that if $\hat{g}$ is a solution of this equation then $\hat{g}^2$ is also a
solution, which implies
\begin{equation}
 \hat{g}^2 = \hat{a}_0 + a_1 \hat{g},
\end{equation}
where $\hat{a}_0$ is a constant matrix and $a_1$ is a complex number.
It can be further shown that the product of any two solutions of the Eilenberger
equations is also a solution of these equations. This gives a very
powerful relation,
\begin{equation}
\hat{g}^2 = \hat{a}_0=\hat{g}^2_{bulk},\label{eq:nrm}
\end{equation}
which is very useful in obtaining the numerical solutions of these equations.
There are multiple solutions to this system of equations. In pure superconducting
state, there are three independent solutions. There are two divergent solutions
along with a bounded physical solution. All the unphysical solutions decay to zero
in the bulk. Let's consider two such unphysical solutions, $\hat{g}_{\pm}\propto
e^{\pm \nu \lambda}$. It can be shown that commutator of these two unphysical
solutions gives the physically bounded solution,
\begin{eqnarray}
 \hat{X}_{\pm} &=& \hat{g}_- \hat{g}_+ \pm \hat{g}_+ \hat{g}_-,  \label{eq:xpm} \\
 \hat{\dot{X}}_\pm&=& [\mathcal{T},\hat{X}_\pm].
\end{eqnarray}
The bounded physical solution is,
\begin{eqnarray}
 \hat{g}_p = \hat{c}_p \hat{X}_{-},
 \label{eq:phys}
\end{eqnarray}
the constant $\hat{c}_p$ can be determined using Eq. \eqref{eq:nrm} and \eqref{eq:phys}
and it reads,
\begin{equation}
 \hat{c}_p = \frac{1}{\hat{X}_+}.
\end{equation}
We use these unphysical solutions $\hat{g}_{+,-}$ in the bulk  and integrate
towards the vortex core starting from the bulk. Since these solutions grow
exponentially,  they can be easily computed numerically with appropriate
boundary conditions. Far away from defects, we can ignore the spatial dependence
of the order parameters. Since there is no long-range SDW order, we have the
standard Eilenberger equations for the pure superconducting state in the  bulk
and in the basis we consider here, the Eilenberger Green's function is a block
diagonal matrix. The two bands are coupled through the self-consistency
condition. Therefore it is sufficient to illustrate the idea for one band, for
which we write down the equations explicitly,
\begin{eqnarray}
 \dot{g} &=& i\left( \Delta^* f + \Delta f^\dagger \right), \\
 \dot{f} &=& -2\omega f - 2i \Delta g, \\
 \dot{f}^{\dagger} &=& 2\omega f^\dagger -2i\Delta^* g.
\end{eqnarray}
We first find two unphysical solutions, which can be determined easily. Let's
consider,
\begin{eqnarray}
g&=& c_1 e^{\zeta \lambda}, \\
f&=& c_2 e^{\zeta \lambda}, \\
f^\dagger&=& c_3 e^{\zeta \lambda}.
\end{eqnarray}
where $\zeta = \pm \nu$. Normalization condition requires,
\begin{eqnarray}
 \hat{g}^2_\pm = 0.
\end{eqnarray}
This ensures that all the unphysical solution decay to zero in the bulk. Which
gives,
\begin{eqnarray}
 c^2_1+c_2c_3 =0, \\
 c_1 = i\sqrt{c_2 c_3}.
\end{eqnarray}
Using these conditions,
\begin{eqnarray}
 (\zeta + 2\omega)c_2 &=&2\Delta\sqrt{c_2 c_3}, \\
 (\zeta - 2\omega)c_3 &=& 2\Delta^* \sqrt{c_2 c_3}.
\end{eqnarray}
These two equations give the value of $\zeta = \pm 2\sqrt{\omega^2+|\Delta|^2}/v_F$
and
\begin{eqnarray}
 c_2 &= \frac{-2i\Delta}{\zeta v_F+2\omega} c_1, \\
 c_3 &= \frac{-2i\Delta^*}{\zeta v_F-2\omega} c_1.
\end{eqnarray}
Here we fix $c_1 =1$ and write the exploding solutions,
\begin{eqnarray}
 \hat{g}_+ &=& \exp \left[+\frac{2Q \lambda}{v_F}\right] \left[ \begin{array}{cc}
                                     1 & -i\Delta p_+ \\
                                     i\Delta^{*} p_- & -1
                                    \end{array} \right], \\
\hat{g}_- &=& \exp\left[-\frac{2Q\lambda}{v_F}\right] \left[ \begin{array}{cc}
                                     1 & -i\Delta p_- \\
                                     i\Delta^{*} p_+ & -1
                                    \end{array} \right], \\
p_{\pm} &=& \frac{1}{\omega\pm Q}, \\
Q&=& \sqrt{\omega^2+|\Delta|^2}.
\end{eqnarray}
Now we can write down the physical solution,
\begin{eqnarray}
 \hat{g}_p &=& \frac{\hat{X}_-}{\hat{X}_+} =\frac{1}{\sqrt{\omega^2+|\Delta|^2}} \left[ \begin{array}{cc}
                                   \omega & -i\Delta \\
                                   i\Delta^* & -\omega
                                  \end{array} \right]
\end{eqnarray}
Once we get the values of these two diverging solutions in bulk then we can use the
bulk values to integrate towards the vortex core and find the physical solution
using two unphysical solutions.
%
\end{document}